# Low-dimensional gap plasmons for enhanced light-graphene interactions

Yunjung Kim, Sunkyu Yu, and Namkyoo Park*

*Photonic Systems Laboratory, Department of Electrical and Computer Engineering, Seoul National University, Seoul 08826, Korea*

*E-mail address for correspondence: nkpark@snu.ac.kr

**Graphene plasmonics has become a highlighted research area due to the outstanding properties of deep-subwavelength plasmon excitation, long relaxation time, and electro-optical tunability. Although the giant conductivity of a graphene layer enables the low-dimensional confinement of light, the atomic scale of the layer thickness is severely mismatched with optical mode sizes, which impedes the efficient tuning of graphene plasmon modes from the degraded light-graphene overlap. Inspired by gap plasmon modes in noble metals, here we propose low-dimensional graphene gap plasmon waves for large light-graphene overlap factor. We show that gap plasmon waves exhibit superior in-plane and out-of-plane field concentrations on graphene compared to those of edge or wire-like graphene plasmons. By adjusting the chemical property of the graphene layer, efficient and linear modulation of graphene gap plasmon modes is also achieved. Our results provide potential opportunities to low-dimensional graphene plasmonic devices with strong tunability.**

## Introduction

In the context of light-matter interactions, the concentration of electromagnetic fields on materials is a critical issue for the performance of tunable optical devices, such as photodetectors[1], bio-sensors[2], optical modulators[3,4], and lasers[5]. Plasmonic structures[6-9] have thus been intensively studied to achieve subwavelength field concentration. For the design of plasmonic devices, the proper selection of metals determines the boundary of device performances[10] for power consumption, bandwidth, and footprints.

Due to its two-dimensional (2D) structure with extremely large conductivity from the massless Dirac point[11,12], graphene has become a leading candidate for deep-subwavelength plasmonics[13-15]. Along with its structural advantage for the integration, the giant and tunable conductivity of the graphene layer also enables the modulation of its optical properties. A number of devices such as absorber[16,17], modulators[18,19], and tunable metamaterials[20-22] controlling optical flows through the designed graphene layer have been proposed and demonstrated, by manipulating the dispersion of graphene conductivity via electric gating[20,23-25] or chemical doping[26,27]. However, the atomically-thin graphene layer leads to the intrinsic limit for the device performance at the same time; the significant scale mismatch between ~10nm to ~100nm size optical modes and ~Å-scale graphene layers severely degrades the light-graphene overlap which prohibits the efficient manipulation of light flows. The achievement of small modal size[28,29] and more importantly, the high overlap factor with the graphene layer, is thus an urgent issue for tunable graphene plasmonics.

Here, we focus on low-dimensional waveguide systems for the superior light-graphene overlap factor. We firstly reveal the existence of graphene gap plasmon (GGP) modes the field profile of which is strongly confined inside the graphene gap between metallic and dielectric graphene layers. We demonstrate that the GGP mode has larger field concentration on graphene layers than those of edge[30,31] or wire-like graphene plasmon modes[28]. By exploiting the tunable graphene conductivity through the chemical potential modulation, highly sensitive and linear modulation of the GGP propagation constant is also achieved with its stable mode profile. The proposed low-dimensional waveguide systems with the improved overlap factor pave the path toward integrated plasmonic devices on graphene.

**Results**

We consider the 2D metal-gap-dielectric waveguide system composed of three distinct graphene domains (Fig. 1a); a dielectric gap domain G with the width $w$ (the sheet conductivity $\sigma^{(G)}$ where $Im\{\sigma^{(G)}\} < 0$) is inserted in-between semi-infinite metallic domain M ($Im\{\sigma^{(M)}\} > 0$) and dielectric domain D ($Im\{\sigma^{(D)}\} < 0$). The gap region G satisfies the condition of $Im\{\sigma^{(D)}\} < Im\{\sigma^{(G)}\} < 0$, as the analogy of plasmonic gap modes in noble metals[8,9]. Note that such a system can be achieved by applying the spatial variation of the conductivity on a single graphene layer, based on the tuning of its chemical potential[20,23-25] (or doping level) $\mu(x)$. Figure 1b shows the $\mu$-dependency of the graphene conductivity calculated by Kubo formula[11,20,22], which determines the operation regime for each layer (frequency $f = \omega / 2\pi = 20$ THz, charged particle scattering rate[11] $\Gamma = 0.43$ meV, and temperature T = 3 K). To satisfy the gap mode condition as similar to the case of noble metals[8,9], the proposed system can be realized by adjusting the doping level of each graphene region corresponding to $(\Omega^{(M)})^{-1} > 0.6$ and $0.5 < (\Omega^{(D)})^{-1} < (\Omega^{(G)})^{-1} < 0.6$, where $\Omega^{-1} = \mu/(\hbar\omega)$ is the normalized chemical potential.

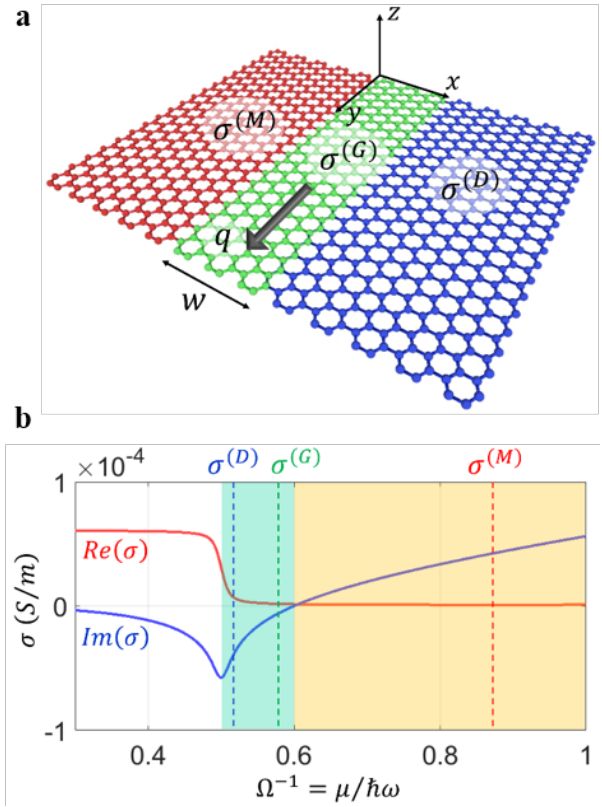

**Figure 1. Low-dimensional (2D) graphene waveguide system for GGP modes, consisting of three distinct graphene domains.** (**a**) A schematic of the proposed 2D metal-gap-dielectric system, composed of a gap domain G (dielectric, $0 < x < w$; $\sigma^{(G)}$) between two semi-infinite domains M (metallic, $x < 0$; $\sigma^{(M)}$) and D (dielectric, $x > w$; $\sigma^{(D)}$). The GGP mode is assumed to propagate along the *y*-axis with the wavevector $\boldsymbol{q} = q\boldsymbol{k}$. (**b**) The graphene conductivity as a function of the normalized chemical potential $\Omega^{-1}$, satisfying the condition of $Im\{\sigma^{(D)}\} < Im\{\sigma^{(G)}\} < 0 < Im\{\sigma^{(M)}\}$. The blue-green (or yellow) region denotes the dielectric (or metallic) regime with $0.5 < \Omega^{-1} < 0.6$ (or $\Omega^{-1} > 0.6$). The highly-lossy region ($\Omega^{-1} < 0.5$ with $Re(\sigma) \gg 0$) from the interband transition is excluded in our discussion.

Figure 2a shows the electric field profile of the low-dimensional GGP mode in the 2D metal-gap-dielectric waveguide system, calculated by the eigenmode solver of COMSOL Multiphysics ($w$ = 5nm, $(\Omega^{(M)})^{-1} = 4$, $(\Omega^{(G)})^{-1} = 0.54$, and $(\Omega^{(D)})^{-1} = 0.5002$). In the numerical analysis, the graphene is considered as the film[20] with the thickness[28] of $\delta$ = 0.2nm and the relative bulk permittivity of $\varepsilon_g(\omega) = 1 + j\sigma_g(\omega)$ /

($\omega\varepsilon_0\delta$), where $\sigma_g$ is sheet conductivity of graphene obtained from Kubo formula[11] (Fig. 1b). To demonstrate the distinctive feature of GGP modes, we compare with other graphene waveguide modes: graphene edge plasmon (GEP) mode[30,31] (Fig. 2b) and wire-like 1D-SPP mode[28] (Fig. 2c) with same material parameters. While both GGP and 1D-SPP modes with quasi-antisymmetric potential profiles ($\sigma(x,y) \sim -\sigma(-x,y)$ for all $y$) have superior confinement compared to that of the GEP mode with much stronger structural asymmetry ($|\sigma(x,0)| \ll |\sigma(-x,0)|$ and $|\sigma(x,y)| = |\sigma(-x,y)|$ for $y \neq 0$), GGP exhibits more confined transverse ($E_x$) field on the gap region than that of the 1D-SPP mode, as similar to the difference between gap plasmons and surface plasmons in noble metals[8,9]. This transverse concentration originates from the continuity condition of the displacement current $Im\{\sigma^{(G)}\} \cdot E_x^{(G)} \sim Im\{\sigma^{(D)}\} \cdot E_x^{(D)}$, deriving the enhancement of $E_x^{(G)}$ from the condition of $Im\{\sigma^{(D)}\} < Im\{\sigma^{(G)}\} < 0$.

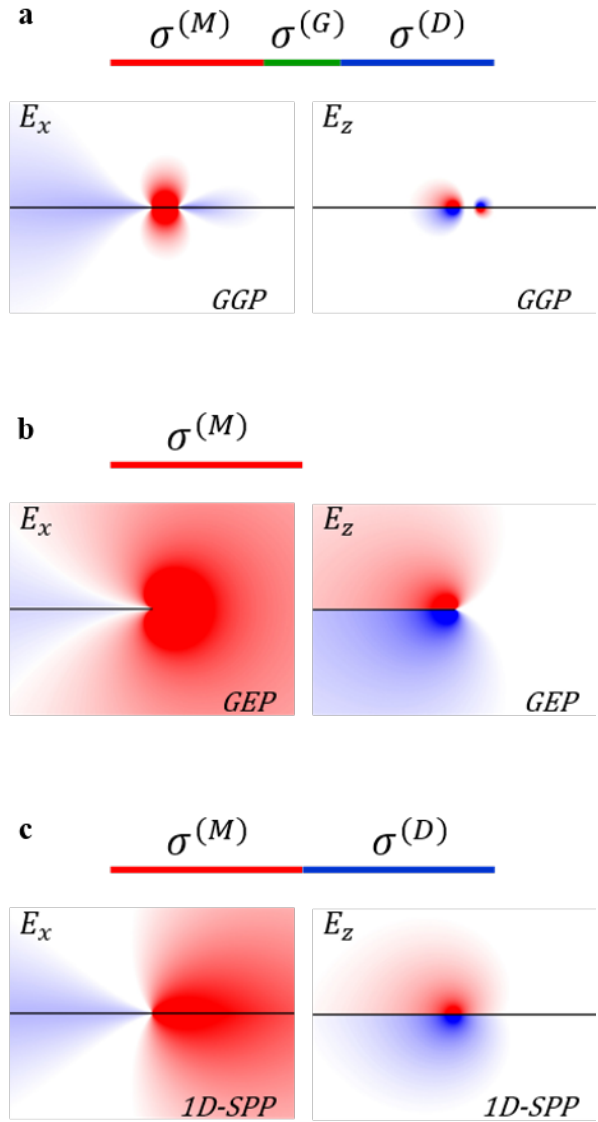

**Figure 2. Electric field distributions of graphene plasmon modes** for (**a**) GGP mode ($w$ = 5nm and $(\varOmega^{(G)})^{-1}$ = 0.54), (**b**) GEP mode, and (**c**) wire-like 1D-SPP mode. $(\varOmega^{(M)})^{-1}$ = 4 for all cases, and $(\varOmega^{(D)})^{-1}$ = 0.5002 for (**a,c**). The horizontal black lines indicate graphene layers.

Figure 3 shows the characteristics of the GGP mode, including effective mode index, intensity profiles, and field concentration. As similar to the case of noble metal gap plasmons[32], many aspects of the GGP mode represent the intermediate features between those of 1D-SPP modes for low and high dielectric graphene layers. For example, the proposed GGP system with zero width ($w = 0$) corresponds to the $\sigma^{(M)}$-$\sigma^{(D)}$ 1D-SPP system with the effective mode index[28] of $n_{\text{eff}} = Re\{q\} / k_0 \approx 2 \cdot (3/2)^{1/2} \cdot \varepsilon \varepsilon_0 \cdot c / (Im\{\sigma^{(M)}\} + Im\{\sigma^{(D)}\})$, while the GGP system with infinite $w$ is converged to the $\sigma^{(M)}$-$\sigma^{(G)}$ 1D-SPP system. The effective mode index (Fig. 3a) and the modal size (Fig. 3b) of the GGP mode with different widths are thus varying between these two boundaries.

Most importantly, there exist differentiated features of the GGP mode when compared to 1D-SPP modes, as shown in Figs 3c and 3d. Figure 3c shows the electric field intensity of the GGP modes along the center of the graphene layer for different gap width (2nm to 30nm). Although the intensity profile of large $w$ case is converged to that of the $\sigma^{(M)}$-$\sigma^{(G)}$ 1D-SPP system, smaller $w$ cases exhibit the intensity profiles focused of the graphene gap. Such a distinct in-plane intensity distribution imposes the unique property on out-of-plane confinement, in terms of the light-graphene overlap factor $\rho = \iint_{\text{graphene}} |E|^2 \cdot dS / \iint |E|^2 \cdot dS$: the concentration of electromagnetic fields on graphene. Figure 3d presents the variation of $\rho$ as a function of the gap width $w$, which demonstrates the superior light-graphene overlap for the structures with apparent field concentration on the gap ($0 < w < 40$nm). We note that the GGP mode acquires much higher field concentration on the graphene layer ($\rho = 2.07 \times 10^{-3}$ at $w = 5$nm), when compared to those of 1D-SPP modes ($\sigma^{(M)}$-$\sigma^{(D)}$ system of $\rho = 1.81 \times 10^{-3}$ and $\sigma^{(M)}$-$\sigma^{(G)}$ system of $\rho = 1.70 \times 10^{-3}$) and the GEP mode ($\rho = 0.728 \times 10^{-3}$).

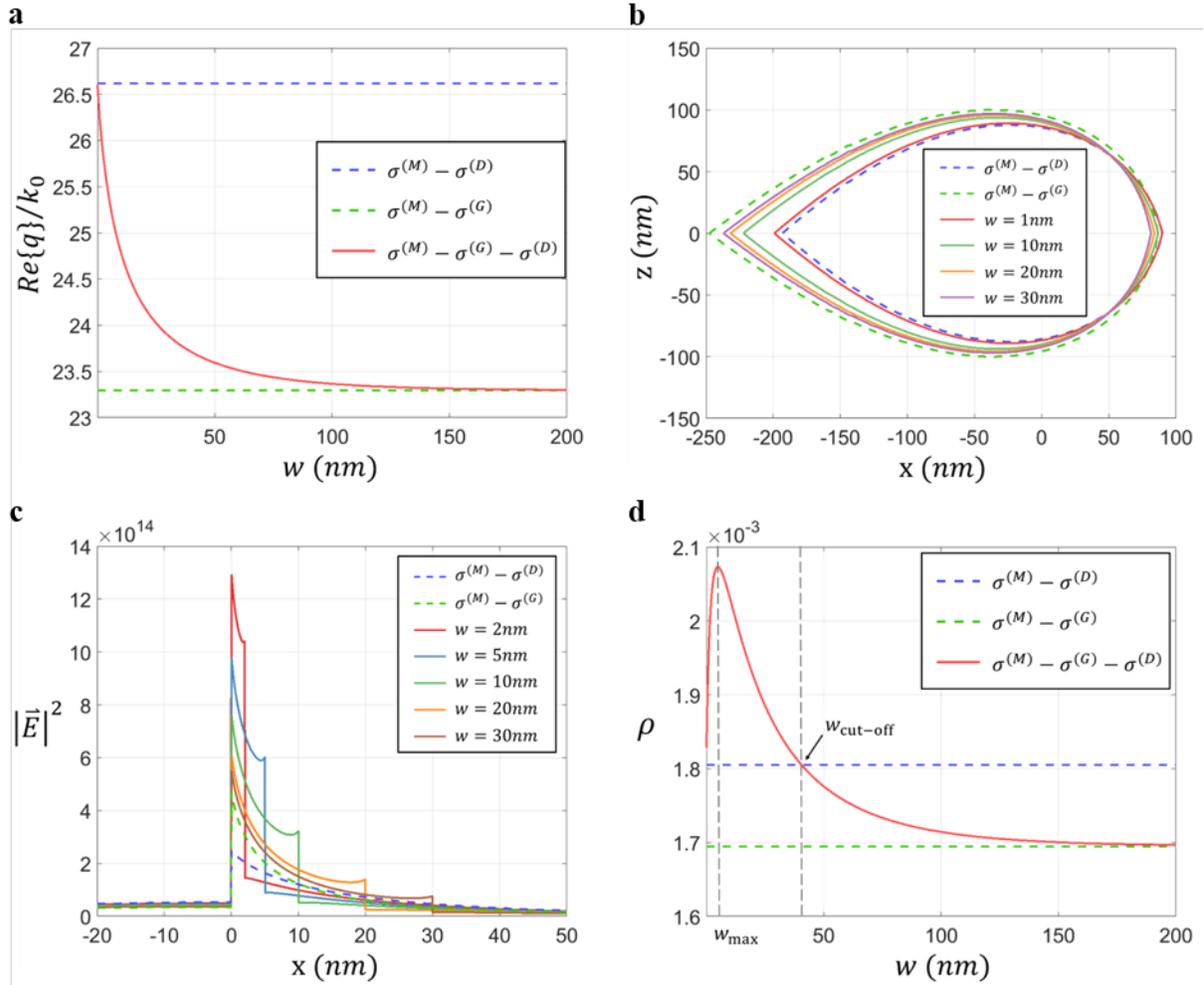

**Figure 3. Modal properties of the GGP mode controlled by the structural parameter *w*.** (**a**) Effective mode index $n_{\text{eff}} = Re\{q\}/k_0$ of the GGP mode as a function of the gap width *w*. (**b**) Modal cross section contours corresponding to *w* = 1, 10, 20, and 30nm, which depict the region A for $\iint_A |E|^2 \cdot dS / \iint |E|^2 \cdot dS = 0.8$. (**c**) Electric field intensity along the center of the graphene layer (*x*-axis) for different gap widths *w*, compared to the cases of $\sigma^{(M)}$-$\sigma^{(D)}$ and $\sigma^{(M)}$-$\sigma^{(G)}$ 1D-SPP systems. (**d**) Graphene field concentration $\rho$ as a function of the gap width *w* ($w_{\text{max}}$ = 5nm, $w_{\text{cut-off}}$ = 40nm). The blue dashed (green dashed) line in (**a-d**) denotes 1D-SPP modes. All other parameters are same as those in Fig. 2.

The large overlap factor in Fig. 3 allows for the enhancement of light-graphene interactions. Figure 4 shows the modulation of GGP modes by controlling the chemical potential of the graphene layer as $(\Omega^{(M)})^{-1} = 4 + \Delta\Omega^{-1}$, $(\Omega^{(G)})^{-1} = 0.54 + \Delta\Omega^{-1}$, and $(\Omega^{(D)})^{-1} = 0.5002 + \Delta\Omega^{-1}$. As seen, the effective mode index of the GGP mode can be controlled with an order of smaller modulation of $\Delta\Omega^{-1}$ when compared to the GEP mode. The GGP mode also provides more efficient regime of $\Delta\Omega^{-1}$ for controlling effective index compared to 1D-SPP modes ($\Delta\Omega^{-1} \leq 0.015$). Such superior efficiency is more apparent for the case of the finite modulation region for $\Delta\Omega^{-1}$ (dotted lines in Fig. 4a, for the $3w_{max}$ modulation width around the graphene gap), due to the superior transverse localization of the GGP mode (Fig. 3c). Note that the spatial profile of electric field intensity (Fig. 4b) and the overlap factor $\rho$ (Fig. 4c) of the GGP mode is highly stable to the change of $\Delta\Omega^{-1}$. This stability allows the adiabatic change of the propagation feature of the GGP mode, which is the origin of the linear variation of $n_{eff}$ versus $\Delta\Omega^{-1}$ in Fig. 4a. Because the change of chemical potentials is usually derived by the external electric field, the sensitive and linear modulation of the GGP mode demonstrated in Fig. 4 enables the high-speed, low-power and distortion-free realization of tunable graphene devices.

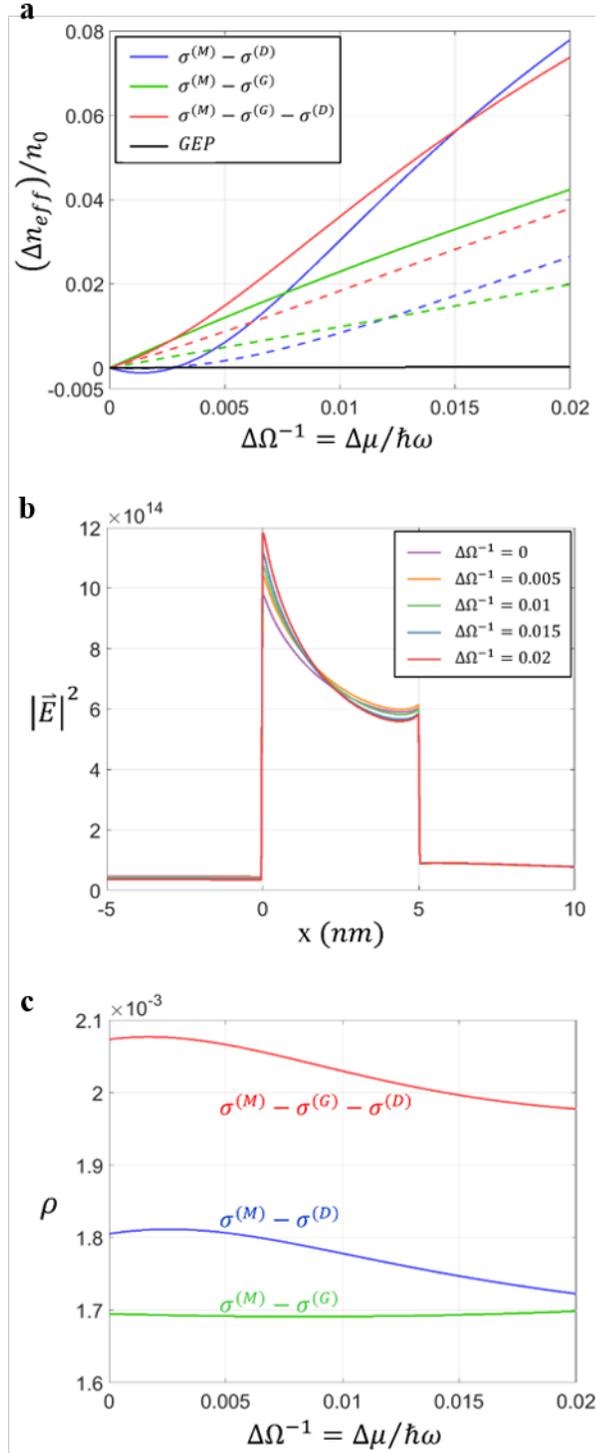

**Figure 4. The effect of the chemical potential modulation on the characteristics of GGP modes. (a)** The variation of effective mode index $n_{eff}$ for GGP, 1D-SPP, and GEP modes for the case of $w_{max}$ = 5nm, as a function of the chemical potential modulation $\Delta\Omega^{-1}$. Dashed lines denote the case of the local

modulation on the transverse plane ($3w_{max}$ modulation width). (**b**) Electric field intensity of the GGP mode along the center of the graphene layer (*x*-axis), for different values of $\varDelta\varOmega^{-1}$. (**c**) Graphene field concentration factor $\rho$ as a function of $\varDelta\varOmega^{-1}$. GEP mode exhibits much lower $\rho$ in (**c**) ($\rho \sim 0.728 \times 10^{-3}$, not shown). All other parameters are same as those in Fig. 2.

**Discussion**

We demonstrated the existence of low-dimensional gap plasmon modes on graphene, which supports large light-graphene overlap factor. The system with spatially-varying chemical potential (or doping level) for GGP modes can be realized by several existing schemes such as electric field bias[20] or substrate level control[33]. Highly efficient manipulation with the stable field profile of the GGP mode, superior to those of GEP[30,31] or wire-like 1D-SPP modes[28], opens the pathway toward tunable graphene plasmonics with high-speed, low-power and distortion-free operation. The modal profile dependency of the light-graphene overlap factor also imposes intriguing opportunity on unconventional wave profiles supported by 2D materials, based on optical transformation techniques[20,34,35].

**Acknowledgments**

This work was supported by the National Research Foundation of Korea (NRF) through the Global Frontier Program (GFP) NRF-2014M3A6B3063708 and the Global Research Laboratory (GRL) Program K20815000003, which are all funded by the Ministry of Science, ICT & Future Planning of the Korean government. S. Yu was also supported by the Basic Science Research Program (2016R1A6A3A04009723) through the NRF, funded by the Ministry of Education of the Korean government.

**Author Contributions**

Y.K. developed the theory and performed eigenmode computations. S.Y. conceived the presented idea for the GGP mode and its light-graphene overlap factor. N.P. encouraged Y.K. and S.Y. to investigate unconventional graphene plasmon modes while supervising the findings of this work. All authors discussed the results and contributed to the final manuscript.

**Competing Interests Statement**

The authors declare that they have no competing financial interests.